# Robust axion insulator and Chern insulator phases in a two-dimensional antiferromagnetic topological insulator

Chang Liu[1,†], Yongchao Wang[2,†], Hao Li[3,4,†], Yang Wu[4,5], Yaoxin Li[1], Jiaheng Li[1], Ke He[1,6], Yong Xu[1,7,*], Jinsong Zhang[1,*], Yayu Wang[1,6*]

[1]*State Key Laboratory of Low Dimensional Quantum Physics, Department of Physics, Tsinghua University, Beijing 100084, P. R. China*

[2]*Beijing Innovation Center for Future Chips, Tsinghua University, Beijing 100084, P. R. China*

[3]*School of Materials Science and Engineering, Tsinghua University, Beijing, 100084, P. R. China*

[4]*Tsinghua-Foxconn Nanotechnology Research Center, Department of Physics, Tsinghua University, Beijing 100084, P. R. China*

[5]*Department of Mechanical Engineering, Tsinghua University, Beijing 100084, P. R. China*

[6]*Collaborative Innovation Center of Quantum Matter, Beijing, P. R. China*

[7]*RIKEN Center for Emergent Matter Science, Wako, Saitama 351-0198, Japan*

† These authors contributed equally to this work.

* Emails: yongxu@tsinghua.edu.cn; jinsongzhang@tsinghua.edu.cn;

yayuwang@tsinghua.edu.cn

The intricate interplay between nontrivial topology and magnetism in two-dimensional (2D) materials has led to the emergence of many novel phenomena and functionalities. An outstanding example is the quantum anomalous Hall (QAH) effect, which was realized in magnetically doped topological insulators (TIs) in the absence of magnetic field[1–6]. Recently, the layered van der Waals compound $MnBi_2Te_4$ has been theoretically predicted and experimentally verified to be a TI with interlayer antiferromagnetic (AFM) order[7–16]. It is a rare stoichiometric material with coexisting topology and magnetism, thus represents a perfect building block for complex topological-magnetic structures. Here we investigate the quantum transport behaviors of both bulk crystal and exfoliated $MnBi_2Te_4$ flakes in a field effect transistor geometry. In the 6 septuple layers (SLs) device tuned into the insulating regime, we observe a large longitudinal resistance and zero Hall plateau, which are characteristic of the axion insulator state. The robust axion insulator state occurs in zero magnetic field, over a wide magnetic field range, and at relatively high temperatures. Moreover, a moderate magnetic field drives a quantum phase transition from the axion insulator phase to a Chern insulator phase with zero longitudinal resistance and quantized Hall resistance $h/e^2$ ($h$ is the Plank constant and $e$ is the elemental charge). These results pave the road for using even-number-SL $MnBi_2Te_4$ to realize the quantized topological magnetoelectric effect and axion electrodynamics in condensed matter systems.

Axion insulator is an exotic topological phase that has zero Chern number but a finite topological Chern-Simons term[17]. It was proposed to be a promising platform for exploring the Majorana edge modes, quantized topological magnetoelectric coupling and axion electrodynamics in condensed matter[17–22]. Previous attempts to construct the axion insulator phase mainly based on fabricating heterostructures of QAH films with different coercive fields[23–25], which require complex epitaxial growth of magnetically doped TIs, and transport measurements at ultralow temperatures and finite magnetic fields. There is an urgent need for finding a stoichiometric material which can achieve robust axion insulator state in zero magnetic field and high temperatures. Interestingly, the recently discovered MnBi$_2$Te$_4$ naturally fulfills such stringent requirements.

The MnBi$_2$Te$_4$ compound has a van der Waals type layered structure with the space group R$\bar{3}$m (Ref. 26). As shown by the schematic crystal structure in Fig. 1a inset, each SL of MnBi$_2$Te$_4$ can be roughly viewed as a quintuple layer of Bi$_2$Te$_3$ TI intercalated with an extra Mn-Te bilayer in the middle. Each Mn$^{2+}$ ion is expected to contribute a 5 $\mu_B$ magnetic moment, and the intralayer exchange coupling gives rise to ferromagnetic (FM) order in each SL with an out-of-plane easy axis[7]. The interlayer exchange coupling between neighboring SLs is AFM, generating the A-type AFM order in three-dimensional bulk[7–16]. The combination of AFM order and nontrivial band topology makes MnBi$_2$Te$_4$ a promising material for finding novel topological phases and phase transitions by either controlling the layer structure or applying magnetic field.

We begin by synthesizing and characterizing the bulk crystal of MnBi$_2$Te$_4$. Figure 1a displays the temperature ($T$) dependence of in-plane resistivity ($\rho_{xx}$) on a MnBi$_2$Te$_4$ crystal with thickness ~ 25 μm, which reveals a sharp transition at the Néel temperature $T_N$ ~ 25 K, consistent with previous reports[12–16]. The magneto-resistivity (MR) curves measured in an out-of-plane magnetic field at varied $T$s are displayed in Fig. 1b. Below $T_N$, $\rho_{xx}$ decreases sharply across a lower critical magnetic field ($H_{c1}$), when the spin order is driven to a canted AFM state[14–16]. As the magnetic field is further increased to an upper critical field ($H_{c2}$), the $\rho_{xx}$ curve forms a kink and then starts to decrease gradually when the spins are polarized to a FM state with suppressed spin fluctuations. The field dependence of Hall resistivity ($\rho_{yx}$) traces are measured at the same $T$s and displayed in Fig. 1c. The overall negative slopes of the Hall traces indicate electron-type bulk charge carriers, possibly due to the existence of antisite defects[13]. From the linear regime at $H < H_{c1}$, we can extract the electron density to be 1.2 ×10$^{20}$ cm$^{-3}$ at $T$ = 1.6 K, and the mobility is ~ 74 cm$^2$/Vs. In the field range $H_{c1} < H < H_{c2}$, the Hall traces exhibit a broad hump, most likely due to an additional topological Hall term arising from non-zero Berry phase in the canted AFM state with non-collinear spin structure[27–30], which also manifests itself as a hollow in $\rho_{xx}$ (Fig. 1b).

Due to the weak van der Waals bonding between adjacent SLs, MnBi$_2$Te$_4$ crystal can be easily cleaved and transferred onto a SiO$_2$/Si substrate, which also serves as a back gate for the field effect transistor device. By using mechanical exfoliation method, we can obtain thin flakes of MnBi$_2$Te$_4$ with lateral size around 10 μm and thickness down to a few SLs, reaching the 2D limit. In this work we focus on exfoliated MnBi$_2$Te$_4$ with thickness of 6 SLs for the following reasons. Firstly, MnBi$_2$Te$_4$ with even number of SLs have been proposed to possess axion insulator phase in zero magnetic field due to the inherent AFM order[7,8], which avoids the technical challenges and magnetic disorders associated with the QAH heterostructures. Secondly, from the electronic structure point of view, 6-SL is sufficiently thick to decouple the top and bottom surface states so as to ensure the nontrivial bulk band topology. Thirdly, 6-SL is thin enough to reduce contribution from extra conduction channels and for the back gate to efficiently deplete the bulk carriers.

We have fabricated and measured two 6-SL MnBi$_2$Te$_4$ field effect transistor devices, and the experimental results are highly consistent with each other. Figure 2a shows the optical image of a 6-SL MnBi$_2$Te$_4$ device used in the measurements, which is fabricated into a Hall bar geometry. Figure 2b shows an atomic force microscope image near the edge of the flake, and the line profile renders a thickness of 8.2 ± 0.1 nm, equivalent to 6.0 ± 0.1 SL of MnBi$_2$Te$_4$ (the thickness of one SL is 1.36 nm). The bulk band structure of 6-SL MnBi$_2$Te$_4$ with AFM order is obtained by first-principles calculations, as shown in Fig. 2c for dispersions along the M–Γ–K directions.

Figure 2d displays the $\rho_{xx}$ vs. $T$ curves of the 6-SL MnBi$_2$Te$_4$ under varied back-gate voltages ($V_g$s), which reveals several interesting features of the system. Firstly, the overall resistivity value has a non-monotonic $V_g$ dependence and reaches a maximum around $V_g$ ~ 30 V. As indicated by the red dashed line in Fig. 2c, at this $V_g$ the Fermi level ($E_F$) is located near the charge neutral point (CNP) within the surface state gap. Secondly, all the curves show a resistivity kink at ~ 20 K, similar to that at $T_N$ = 25 K in bulk MnBi$_2$Te$_4$, suggesting that the $T_N$ is reduced to 20 K for the 6-SL device. Thirdly, the resistivity always becomes insulating at low $T$, in contrast to the metallic behavior of bulk sample. Lastly, the resistivity drop is much more pronounced in the high $V_g$ regime with large electron density, while the $T_N \sim 20$ K is nearly independent of $V_g$. All these observations suggest that the 6-SL sample enters a regime that is distinctively different from the bulk.

To probe the ground state properties of 6-SL MnBi$_2$Te$_4$, we measure the magnetic field dependence of $\rho_{xx}$ and $\rho_{yx}$ at different gate voltages at $T = 1.6$ K. As shown in Fig. 3a, both quantities exhibit highly complex and yet systematic variations. More detailed data set are documented in supplementary Fig. S1, and here we only focus on the key features in three representative regimes. For $V_g \leq 14$ V, $E_F$ lies in the valence band of the surface states and hole-type 2D carriers dominate the transport process, which can be seen from the overall positive slope of the Hall traces. For $V_g \geq 38$ V, in contrast, $E_F$ lies in the surface conduction band and the electron-type 2D carriers give rise to the overall negative slope of the Hall traces. In these two limits, the MR curves show very different behaviors, presumably due to the different surface state electronic structure and scattering mechanisms. Nevertheless, in all MR curves two characteristic magnetic fields can be defined at around 2.5 T and 5.0 T, which may mark the beginning and ending points of the magnetic field driven spin-flipping process in the 6-SL sample. These are interesting topics that deserve thorough investigations, but will not be the main focus of the current work.

The most remarkable features of the 6-SL MnBi$_2$Te$_4$ exist in the gate range 22 V $\leq V_g \leq 30$ V (enclosed by the blue square in Fig. 3a), when $E_F$ lies within the magnetic gap of the surface states so the 2D conduction channels disappear. In this regime, the longitudinal resistivity at weak magnetic field is very large with $\rho_{xx}$ up to 5.5 $h/e^2$, comparable to the value measured at 1 K in the QAH heterostructures[24], and the relatively low $\rho_{xx}$ value at zero magnetic field may be attributed to the existence of AFM domains. More interestingly, the Hall effect remains at zero for an extended field range -3.5 T $< \mu_0 H <$ 3.5 T, forming a wide $\rho_{yx} = 0$ plateau. These are the hallmark experimental signatures of an axion insulator with Chern number $C = 0$, which has been predicted to be the intrinsic ground state of MnBi$_2$Te$_4$ in the AFM state[7,8,11]. With increasing magnetic field strength, $\rho_{xx}$ exhibits a steep decrease between 3.5 T $< |\mu_0 H|$

< 5 T when the magnetic order is transformed from AFM to FM. The residual $\rho_{xx}$ value at $|\mu_0H|$ = 9 T can be as small as 0.018 $h/e^2$ for $V_g$ = 27 V. Meanwhile, $\rho_{yx}$ increases rapidly and approaches the quantized plateau with $\rho_{yx} \sim \pm 0.974$ $h/e^2$ at $|\mu_0H|$ = 9 T. Both behaviors are characteristic of the quantized Hall effect with Chern number $C = \pm 1$ contributed by dissipationless chiral edge state. Similar results of axion insulator and Chern insulator are also observed in another 6-SL thick device (see Supplementary Fig. S2-4 for details).

Figure 3b shows the variation of zero-field $\rho_{xx}$ (red) and low-field $\rho_{yx}$ vs. $H$ slope (blue) with $V_g$. When $E_F$ is tuned within the surface state gap (22 V $\leq V_g \leq$ 34 V), $\rho_{xx}$ exhibits highly insulating behavior, and d$\rho_{yx}$/d$H$ displays a wide zero plateau, which is fundamentally different from the ambi-polar behavior in conventional insulators. The inset illustrates the schematic AFM order and electronic structure of this axion insulator state, in which the surface state is gapped out by opposite magnetization on the top and bottom layers and no gapless edge state is formed. In Fig. 3c we summarize the $V_g$ dependence of $\rho_{xx}$ and $\rho_{yx}$ measured at $\mu_0H$ = -9 T, which clearly shows that $\rho_{yx}$ forms a quantized Hall plateau and $\rho_{xx}$ is strongly suppressed for 20 V $\leq V_g \leq$ 30 V. This well-defined Chern insulator state is distinct from conventional quantum Hall effect because the former one relies on the magnetic gap at the Dirac point, whereas the later one relies on the formation of Landau levels. The Chern insulator state is not an authentic QAH effect neither because a finite magnetic field is needed. The schematic FM order and electronic structure of this Chern insulator state is shown in the inset of Fig. 3c, where the surface states are gapped out by parallel magnetization on the top and bottom layers but a chiral edge state spans across the gap.

To directly visualize the phase diagram of the axion insulator and Chern insulator states in 6-SL MnBi$_2$Te$_4$, in Fig. 3d we summarize the $V_g$ and $H$ dependence of $\rho_{yx}$ measured at $T$ = 1.6 K into a color map. Apparently, there are three distinct regimes in the phase diagram characterized by different Chern numbers. The blue and red regimes correspond to the Chern insulator states with $C = \pm 1$, in which $\rho_{yx}$ is quantized at $+h/e^2$ and $-h/e^2$ respectively. In the central white region resides the robust axion insulator state with $C = 0$, where $\rho_{yx}$ completely vanishes. Notably, the positive $V_g$ of the CNP suggests that the 6-SL sample without gating is hole-doped, which is opposite to the bulk (Fig. 1c). This conclusion can be also seen from the Hall effect measurements discussed above, and may be due to the finite size effect or the hole doping by gas adsorption, the PMMA cover layer and/or plasma-treated substrates (see Method session).

In addition to the $V_g$ dependent transport behavior, we have also studied the $T$ evolution of $\rho_{xx}$ and $\rho_{yx}$ for $V_g$ set to the CNP, as shown in Fig. 4a. This series of data were acquired after the first cool down, when the sample had the best quality because repeated thermal cycling and gate scanning tend to cause gradual degradation. At $T = 1.6$ K, the Hall resistivity reaches a plateau with $\rho_{yx} = 0.984\ h/e^2$ at $\mu_0 H = -9$ T and the residual $\rho_{xx}$ is merely $0.018\ h/e^2$. Even for $T = 5$ K, the $\rho_{yx}$ value at -9 T is still as high as $0.938\ h/e^2$ and the residual $\rho_{xx}$ as low as $0.11\ h/e^2$. Such Hall quantization is quite remarkable given the relatively high $T$ and low electron mobility of the sample. With increasing $T$, the zero Hall plateau in the axion insulator state becomes narrower due to the smaller field scale for AFM to FM transition. It also gradually develops a negative slope due to the thermally activated electron-type charge carriers. The positive MR behavior at low field persists up to 15 K, but becomes nearly indistinguishable for $T \geq T_N \sim 20$ K. A butterfly-like hysteresis is observed in the longitudinal resistivity, similar to that observed in other 2D AFM materials[31,32]. This imperfect cancellation of magnetic moments is probably due to the existence of AFM domains or the asymmetry between AFM coupled magnetic moments[33]. In Fig. 4b we display the continuous $T$ evolution of $\rho_{xx}$ (red) and $\rho_{yx}$ (blue) for the Chern insulator state at $\mu_0 H = -9$ T. With decreasing $T$, $\rho_{xx}$ decreases along with the sharp increase of $\rho_{yx}$ towards $h/e^2$. The $\rho_{xx}$ curve cannot be fitted by simple thermal activation behavior, indicating the absence of a fixed gap because the magnetization-induced exchange gap varies with $T$.

To examine how the axion insulator to Chern insulator transition takes place, we carried out more detailed measurements at low $T$s. Figure 4c shows that all the $\rho_{xx}$ vs. $H$ curves taken between $T = 1.6$ K and 3 K cross at the same magnetic field $\mu_0 H_c = 4.55$ T. For magnetic field just below $H_c$, $\rho_{xx}$ increases with decreasing $T$, corresponding to an insulating ground state. For magnetic field just above $H_c$, $\rho_{xx}$ decreases with lowering $T$, reminiscent of the metallic ground state of a quantum Hall liquid[34–36]. In order to give a quantitative insight about the quantum phase transition, we perform the scaling analysis of $\rho_{xx}$ with respect to $T$ and $H$ in the vicinity of $H_c$. The $T$ range for scaling is selected according to the widely used criteria that all the $\rho_{xx}$ curves cross a fixed point[37]. As shown in Fig. 4d, all the $\rho_{xx}$ vs. $(H-H_c)/T^{1/\kappa}$ traces collapse into a single curve with $\kappa \sim 0.47$ and critical resistivity $\sim 0.87$ $h/e^2$. Both the two scaling parameters are very close to that of the conventional quantum Hall phase transition[34–36], suggesting that the axion insulator to Chern insulator transition may belong to the same universality class. Figure 4e displays the phase diagram of the 6-SL MnBi$_2$Te$_4$ as a function of $T$ and $H$. The black dashed line marks the boundary between the axion and Chern insulator phases, which is extracted from the $\rho_{yx}$ value at the critical point. The wide magnetic field range and high onset $T$ indicate the robustness of the axion insulator state. At the lowest $T$, the axion insulator state is driven into the Chern insulators state by magnetic field as marked by red color.

The magnetic field driven quantum phase transition from axion insulator to Chern insulator observed here demonstrates the intimate relationship between topology and magnetism in TI, which can be used to toggle between two distinct topological phases. The realization of such exotic states of matter at relatively high temperature in a simple stoichiometric compound makes the 6-SL MnBi$_2$Te$_4$ a perfect material platform for exploring other emergent topological quantum phenomena. In particular, the robust zero-magnetic-field axion insulator state in an easily accessible material significantly facilitates the search for quantized topological magnetoelectric/optical effects and axion electrodynamics in condensed matter systems[18–22].

*Note added*: During the preparation of this manuscript, we became aware of a recent report on the transport properties of exfoliated MnBi$_2$Te$_4$ with odd numbers of SLs[38].

**Methods**

**Crystal growth** High-quality MnBi$_2$Te$_4$ single crystals were grown by direct reaction of 1:1 mixture of high-purity Bi$_2$Te$_3$ and MnTe in a sealed silica ampoule under a dynamic vacuum. The mixture was first heated to 973 K then slowly cooled down to 864 K, and the crystallization occurred during the prolonged annealing at this temperature. The phase and quality of the crystals were examined on a PANalytical Empyrean diffractometer with Cu Kα radiation.

**Device fabrication** Thin MnBi$_2$Te$_4$ flakes were exfoliated by using the Scotch tape method onto 285 nm-thick SiO$_2$/Si substrates, which were pre-cleaned in air plasma for 5 minutes with ~ 125 Pa pressure. Before spin coating of Poly(methyl methacrylate) (PMMA), the surrounding thick flakes were removed by a sharp needle. By using electron-beam lithography, a marker array was first prepared on the substrates for precise alignment between selected flake and patterned electrodes. All the device fabrication processes were carried out in argon-filled glove box with the O$_2$ and H$_2$O levels below 0.1 PPM. Metal electrodes (Cr/Au, 5/50 nm) were deposited in a thermal evaporator connected to the glove box. When transferred between glove box, electron-beam lithography, and the cryostat, the devices were always covered by a layer of PMMA to mitigate air contamination and sample degradation. We have fabricated and measured two 6-SL devices, denoted as Device A and B respectively. The data shown in the main figures are all taken from Device A, and that of Device B are displayed in supplementary Fig. S3 to S5.

**Transport measurement** Electrical transport measurements were carried out in a cryostat with base temperature ~ 1.6 K and magnetic field up to 9 T. The longitudinal and Hall voltages were detected simultaneously by using lock-in amplifiers with AC current generated by a Keithley 6221 current source. The back-gate voltages were applied by a Keithley 2400 source meter.

**Theoretical calculation** First-principles calculations were performed by density functional theory (DFT) using the Vienna *ab initio* Simulation Package[39]. The plane-wave basis with an energy cutoff of 350 eV, the Monkhorst-Pack **k**-point mesh of 7×7×1, the projector augmented wave potential[40] and the Perdew-Burke-Ernzerhof (PBE) exchange-correlation functional in the generalized gradient approximation (GGA) were used[41]. The GGA+$U$ method was applied to describe the localized 3$d$ orbitals of Mn atoms and $U$ = 4.0 eV was selected according to previous tests[7]. The structural relaxation of both lattice constants and atomic positions was performed using with a force criterion of 0.01eV/Å, for which van der Waals corrections were included by the DFT-D3 method[42]. Spin-orbit coupling was included in self-consistent electronic-structure calculations.

**Data availability:** All raw and derived data used to support the findings of this work are available from the authors on request.

**Acknowledgements:** We thank Wenhui Duan, Shoushan Fan, Xiao Feng, Zhenqi Hao, Lexian Yang, and Shusen Ye for helpful discussions and technical supports. This work is supported by the Basic Science Center Project of NSFC (grant No. 51788104), National Key R&D Program of China (grants No. 2018YFA0307100 and No. 2018YFA0305603), and MOST of China (grant No. 2015CB921000). This work is supported in part by the Beijing Advanced Innovation Center for Future Chip (ICFC).

**Author contributions:** Y. Y. W., J. S. Z. and Y. X. proposed and supervised the research. C. L. and Y. X. L. carried out the transport measurements. Y. C. W. fabricated and characterized the devices. H. L. and Y. W. grew the MnBi$_2$Te$_4$ bulk crystals. Y. X. and J. H. L. performed first-principles calculations. Y. Y. W., J. S. Z., Y. X. and C. L. prepared the manuscript with comments from all authors.

**Figure captions**

**Fig. 1 | Transport characterization of MnBi$_2$Te$_4$ single crystal. a**, Temperature dependence of $\rho_{xx}$ measured from room temperature to 1.6 K. The sharp AFM transition is revealed near $T_N$ = 25 K. Inset: schematic crystal structure of MnBi$_2$Te$_4$. The red/blue arrows indicate the magnetic moment directions of Mn ions. **b**, MR curves in an out-of-plane magnetic field at varied temperatures. The lower and upper critical fields ($H_{c1}$, $H_{c2}$) are labeled by black dashed lines. The temperatures for MR curves are shown by the color gradient from red (25 K) to blue (1.6 K). **c**, Magnetic field dependence of Hall resistivity traces measured at the same temperatures. An offset of 5 μΩ.cm is applied.

**Fig. 2 | Characterization of a 6-SL MnBi$_2$Te$_4$ device. a**, Optical image of a 6-SL MnBi$_2$Te$_4$ device fabricated on SiO$_2$/Si substrate. **b**, Atom force microscope imaging close to the flake edge. The line cut reveals a height difference of 8.2 ± 0.1 nm, corresponding to 6.0 ± 0.1 SLs. **c**, Band structure of 6-SL MnBi$_2$Te$_4$ with AFM ordering for dispersions along the M–Γ–K directions. The approximate positions of $E_F$ for $V_g$ = 14, 30 and 50 V are indicated by red dashed lines. **d**, Temperature dependence of $\rho_{xx}$ under different back-gate voltages. All the curves show an overall insulating behavior with a resistivity kink at 20 K. It can be identified as the $T_N$ of the 6-SL sample, and it is nearly independent of $V_g$.

**Fig. 3 | Gate dependent transport properties and magnetic-field-driven axion insulator to Chern insulator transition. a**, Magnetic-field dependence of $\rho_{xx}$ and $\rho_{yx}$ at different gate voltages at $T$ = 1.6 K. When $E_F$ lies within the surface band gap for 22 V ≤ $V_g$ ≤ 30 V (blue square envelope), both the large longitudinal resistivity and wide zero Hall plateau are key signatures of axion insulator state. At high magnetic field, the nearly quantized Hall plateau and vanishing $\rho_{xx}$ are characteristics of a Chern insulator. The red and blue traces represent magnetic field sweeping to the negative and positive directions. **b**, The $V_g$ dependence of $\rho_{xx}$ and the slope of $\rho_{yx}$ vs. $H$ measured at $T$ = 1.6 K around zero magnetic field. The slopes of $\rho_{yx}$ are obtained by linear fitting between ± 2 T, which shows that the carrier density changes from 8.0×10$^{11}$ cm$^{-2}$ for holes at $V_g$

= 0 to $1.3 \times 10^{12}$ cm$^{-2}$ for electrons at $V_g$ = 50 V. **c**, The evolution of $\rho_{xx}$ and $\rho_{yx}$ as a function $V_g$ at $T$ = 1.6 K and $\mu_0 H$ = -9 T, which reveals the well-defined Chern insulator state. The schematic illustration of the magnetic order and electronic structure of the axion insulator and Chern insulator states are displayed in the insets of **b** and **c**. **d**, Experimental phase diagram obtained by a colored contour plot of the $\rho_{yx}$ value at different $V_g$s and magnetic fields at $T$ = 1.6 K. The axion insulator ($C$ = 0) and Chern insulator ($C$ = ±1) states can be directly visualized.

**Fig. 4 | Temperature evolution and quantum critical behavior of the axion insulator to Chern insulator transition. a**, Magnetic-field dependence of $\rho_{xx}$ and $\rho_{yx}$ at varied temperatures with $V_g$ = 25 V. At the base temperature ($T$ = 1.6 K) and high magnetic field ($\mu_0 H$ = -9 T), the $\rho_{yx}$ plateau reaches 0.984 $h/e^2$ and $\rho_{xx}$ is merely 0.018 $h/e^2$. At elevated temperatures, the zero Hall plateau becomes tilted with negative slope at low field. **b**, $T$ dependence of $\rho_{xx}$ and $\rho_{yx}$ for the Chern insulator state ($\mu_0 H$ = -9 T). **c**, The $\rho_{xx}$ vs. $H$ curves in the vicinity of critical field $H_c$ (black arrow) for $T$ = 1.6, 2, 2.5 and 3 K. At the low/high field side of $H_c$, $\rho_{xx}$ increases/decreases with lowering temperature, corresponding to the insulating/metallic behavior reminiscent of the quantum Hall insulator/liquid states. **d**, The scaling analysis of $\rho_{xx}$ in the vicinity of $H_c$. The scaling parameter $\kappa \sim 0.47$ is obtained by fitting $\rho_{xx}$ vs. $(H-H_c)/T^{1/\kappa}$ profiles, which collapse into one single curve with critical resistivity $\sim 0.87$ $h/e^2$. **e**, Experimental phase diagram of 6-SL MnBi$_2$Te$_4$ as a function of $T$ and $H$ at $V_g$ = 25 V. The value of $\rho_{yx}$ is represented by the color map with the color scale shown on the right. Black dashed line marks the phase boundary between the axion insulator and Chern insulator states.

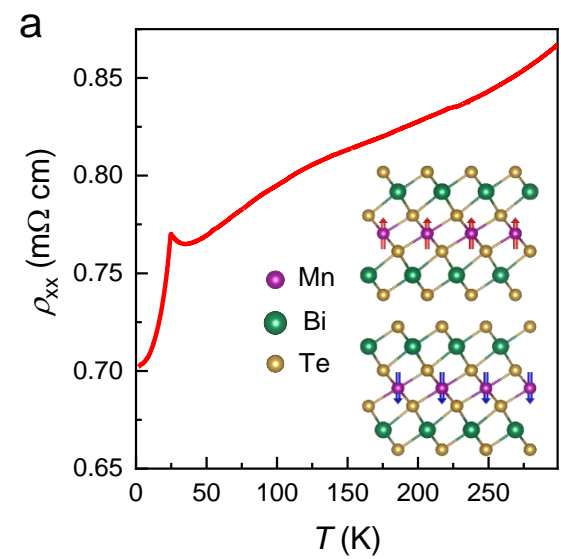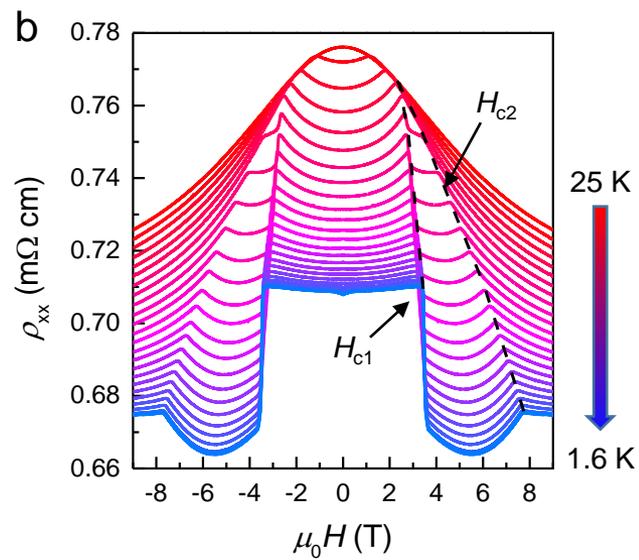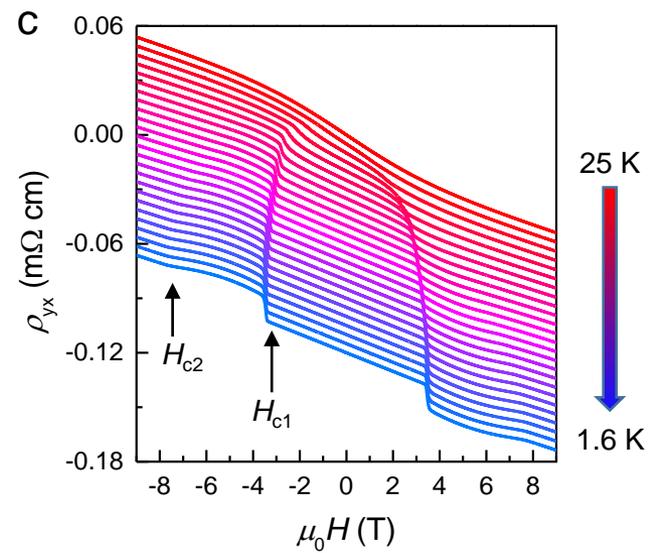

Fig. 1

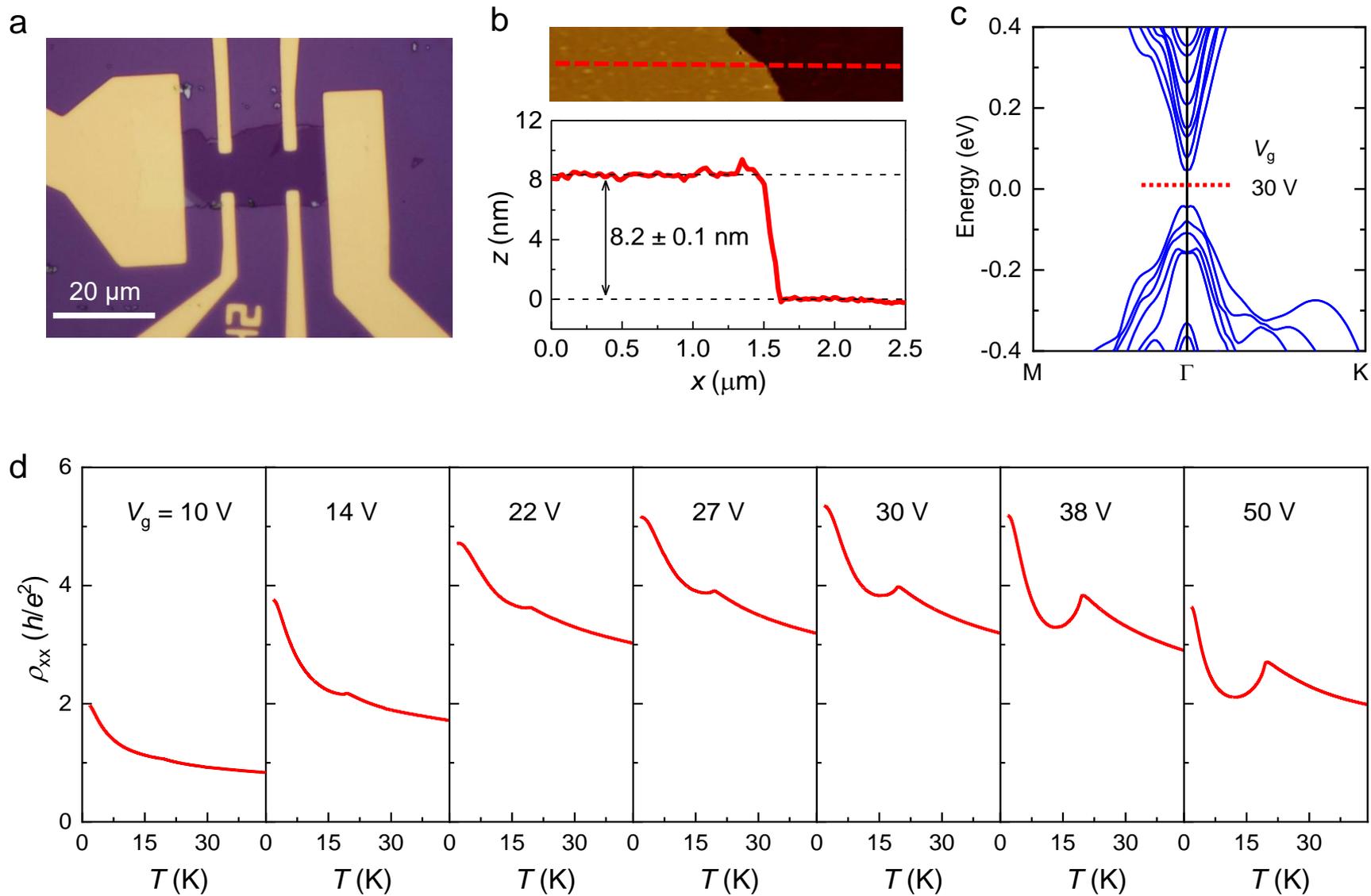

Fig. 2

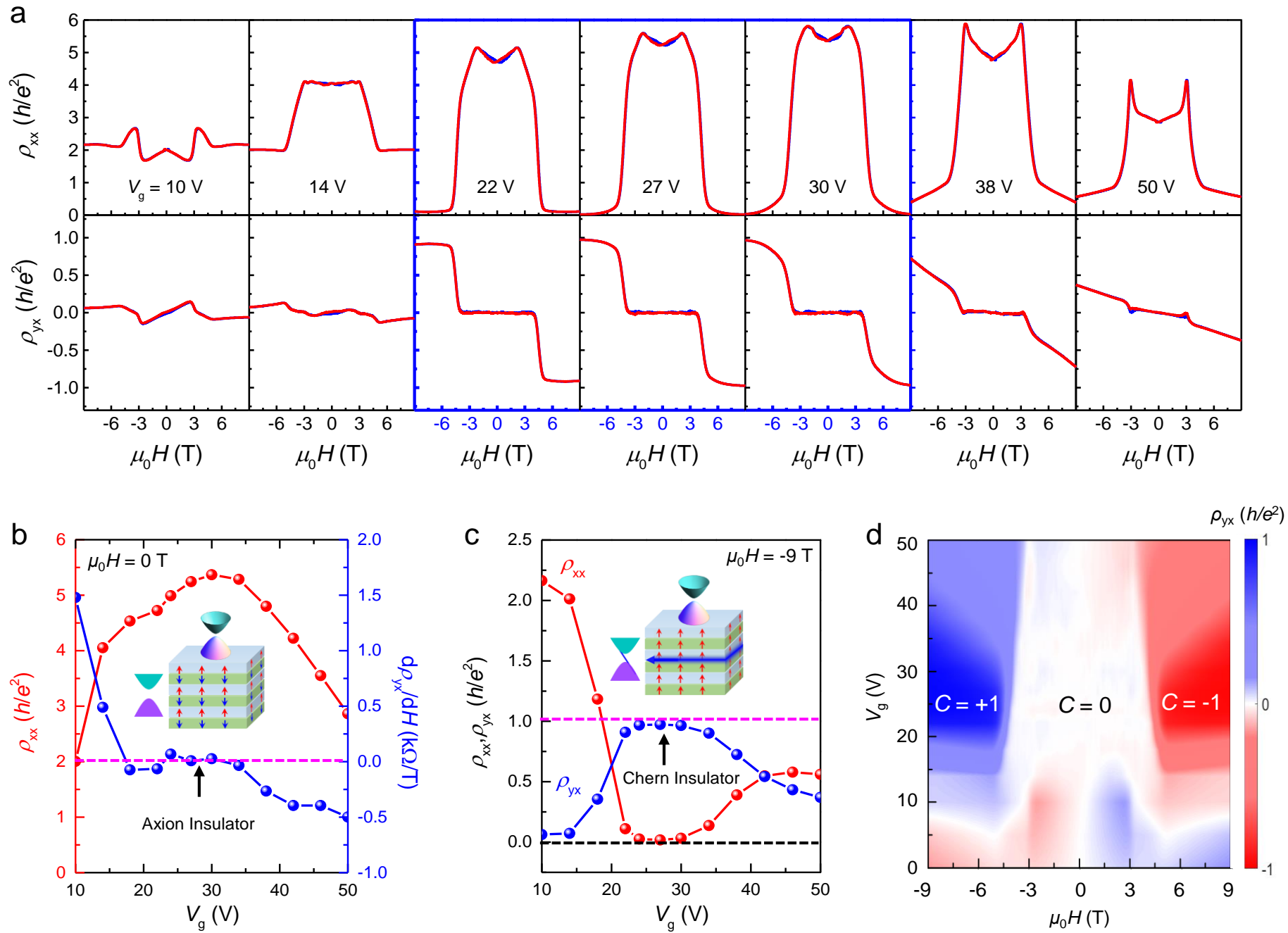

Fig. 3

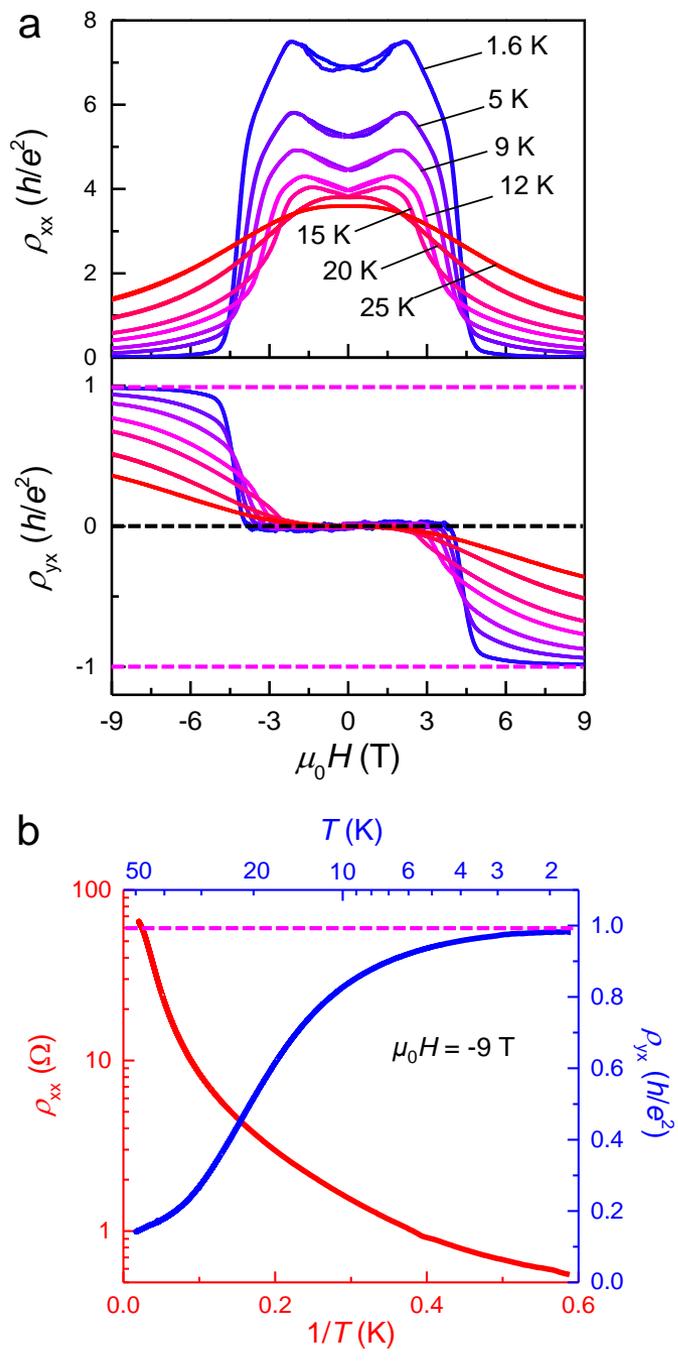
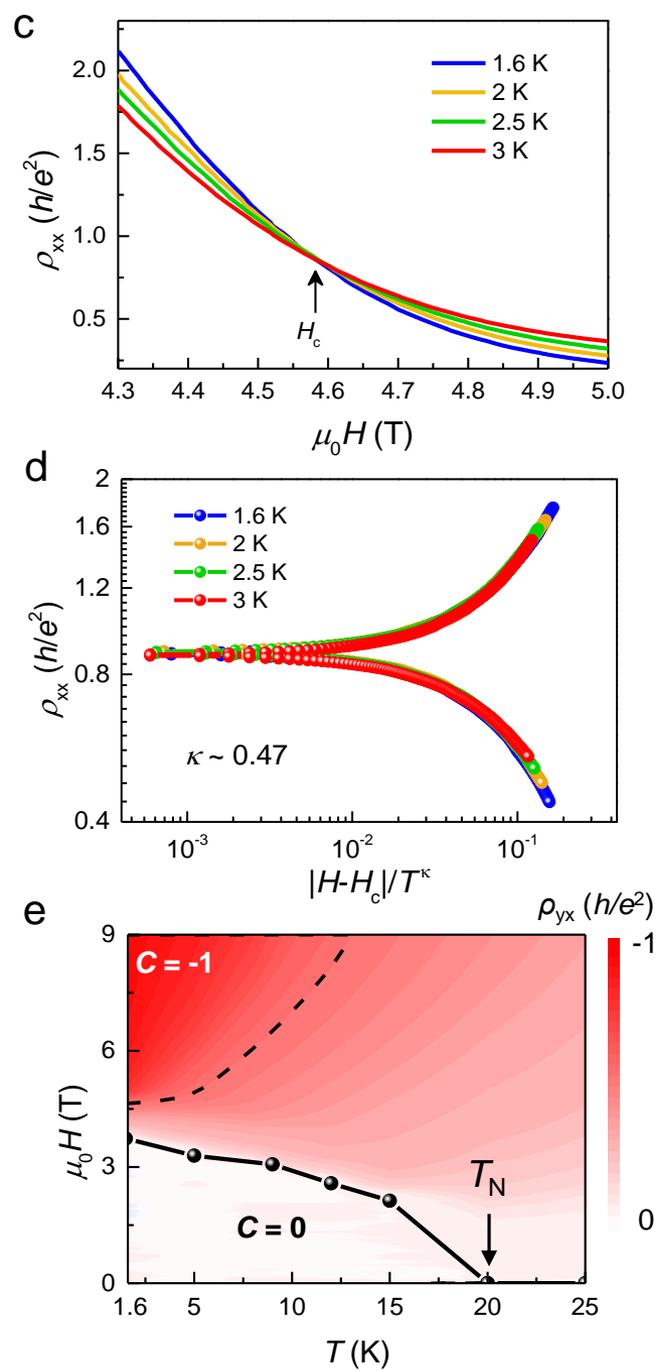

Fig. 4